\documentclass[aps,preprint,showkeys]{revtex4}

\usepackage{graphics}
\usepackage{graphicx}
\usepackage{dcolumn}
\usepackage{bm}

\begin{document}

\title{A Test of the Calculability of a Three Body Relativistic, Cluster Decomposable, Unitary, Covariant Scattering Theory}

\author{Marcus Alfred} 

\author{James Lindesay}

\affiliation{Computational Physics Laboratory\\ Howard University, Washington, D.C. 20059}

\date{\today}
                             
\begin{abstract}
In this work a calculation of the cluster decomposable formalism \cite{lindesay1} for relativistic
scattering as developed by Lindesay, Markevich, Noyes, and Pastrana (LMNP) is made
for an ultra-light quantum model.  After highlighting areas of the theory vital for
calculation, a description is made of the process to go from the general theory to an eigen-integral equation for bound state problems, and calculability is demonstrated.  An ultra-light quantum exchange model is then developed to examine calculability.
\end{abstract}


\maketitle

\section{\label{sec:level1}Introduction}

Several models in nuclear and particle physics have recently demonstrated the need to
expand our understanding of the nature of these systems \cite{carbonell1}.  For example, some have
used the Bethe-Salpeter formalism \cite{salpeter1}, while others use that of Blankenbecler-Sugar \cite{blankenbecler1}. 
Unfortunately in both cases, full cluster decomposability is not satisfied due to Lorentz
kinematics and off shell momentum conservation. In the approach taken in this paper, a
model is developed that satisfies clustering, includes both Lorentz kinematics and
momentum conservation, and finally, is unitary.  
     
We begin by outlining and briefly describing the contents of the sections of this paper.  In
the introductory section, we will summarize the formalism of Lindesay et al \cite{lindesay1} (LMNP)
and their explicit generalization \cite{alfred1}.  Next after introducing relevant parameters and
definitions, we will derive and state the results of a light particle model and its ultralight
extension. 

\section{\label{sec:level2}Background}
    
In this section we will briefly summarize some relevant points of \cite{lindesay1}.  The
formalism yields Lipmann-Schwinger like equations with Faddeev method
decomposition.

\begin{equation}
W_{ab}(z)=-T_{a}(z)R_{0}(z)T_{b}(z)-\sum_{c}\overline{\delta }_{ab}T_{a}R_{0}(z)W_{cb}(z)
\end{equation}

In the preceeding expression, $z$ is the three body off shell parameter, $T_a$ and $T_b$ are two body
$T$ matrices and $R_0$ the resolvent, $W_{ab}$ is the Faddeev component of the fully connected
transition amplitude. 
     
As in traditional non-relativistic, three particle scattering formalisms, LMNP incorporates
two particle kinematics into the three particle kinematics.  This is accomplished by the
introduction of two body off shell and invariant energy terms $\zeta_{(a)}$ and $W_{(a)}$.

\begin{eqnarray}
\zeta _{(a)} & = & \frac{z-e_{a}}{u_{\left( a\right) }^{0}} \\*
W_{\left( a\right)} & = & \frac{M-e_{a}}{u_{\left( a\right) }^{0}}
\end{eqnarray}

In the above equation, $e_a$ is the spectator asymptotic energy, $u^0_{(a)}$ is the zeroth component     for the pair four-velocity, and $M$ is the three body invariant energy.
   
A significant difference between the non-relativistic and LMNP formalisms is the     requirement by the latter of conservation of velocity rather than conservation of
momentum by the former.  This is required if it is desired to have all pairs of virtual     transitions to occur in the same inertial reference frame.  It is found that momentum and
energy conservation are recovered on the energy shell.  In addition, LMNP, like     traditional non-relativistic formalisms, does cluster decompose.  With the inclusion of
Faddeev-like decompositions, it is found that by embedding two particle dynamics into     three particle space, clustering is recovered.
   
We now introduce the two body t matrix form as shown in \cite{alfred1} and used within the models     of this paper,

\begin{eqnarray}
\left\langle k_{a+}k_{a-}\left| t_{a}(z)\right| k_{a+0}k_{a-0}\right\rangle \equiv \left( u_{(a)}^{0}\right) \left( M_{(a)}^{\prime }M_{(a)}\right) ^{-\frac{3}{2}} \nonumber\\
\times \left( \frac{m_{a+}m_{a-}}{\left| \underline{P}_{(a)}\right| \left| \underline{P}_{(a)}^{\prime }\right| }\right) ^{-\frac{1}{2}} 
\delta ^{3}\left( u_{(a)}-u_{(a)}^{\prime }\right) 
\theta \left( W_{(a)}-\left( m_{a+}+m_{a-}\right) \right) \nonumber\\* 
\times \theta \left( W_{(a)}^{\prime }-\left( m_{a}+m_{a-}\right) \right) 
\tau _{a}\left( M_{(a)}\widehat{P}_{(a)}|M^{\prime }_{(a)0}\widehat{P}_{(a)0}^{\prime };z\right)
\end{eqnarray}
$M_{(a)}$ is the pair energy for particles $a_+$ and $a_-$, $u_{(a)}$ is the pair velocity and $k_a$, $k_{a+}$, $k_{a-}$ are
momenta.  The theta functions are included for proper integration, while $\tau_{(a)}(z)$ is the two
body scattering amplitude, and $P_{(a)}$ is the $a_-$  particle momentum in the pair frame.
     
For the three body T matrix, $\rho$  is the form of the Jacobian for the change of variables [5]
from standard particle momenta to invariant energies and velocities.

\begin{eqnarray}
\left\langle k_{1}k_{2}k_{3}\left| T_{a}(z)\right| k^{\prime }_{1}k^{\prime }_{2}k^{\prime }_{3}\right\rangle 
=
\left( u_{(a)}^{0}\right) \left( u^{0}\right) \nonumber\\
\times \left[ \rho _{a}\left( W_{(a)}u^{0}_{(a)}\right) \rho _{d}\left( W^{\prime }_{(a)}u^{0}_{(a)}\right) \right] ^{-\frac{1}{2}}
\delta ^{3}\left( \underline{u}-\underline{u}^{\prime }\right) 
\nonumber\\
\times \delta ^{3}\left( \underline{u}_{(a)}-\underline{u}_{(a)}^{\prime }\right)
\theta \left( W_{(a)}- e_{a} - \left( m_{a+}+m_{a-}\right) \right) 
\nonumber\\
\times \theta \left( W_{(a)}^{\prime }- e_{a} -\left( m_{a}+m_{a-}\right) \right)
\tau _{a}\left( W_{(a)}\widehat{P}_{(a)}|W^{\prime }_{(a)}\widehat{P}_{(a)}^{\prime }; \zeta_{(a)} \right)
\end{eqnarray}

The unitarity of the $W_{ab}$ components of all models is then guaranteed by Friedman,
Lovelace, Namyslowski \cite{freedman1} if the following is true:

\begin{eqnarray}
\tau \left( W_{(a)},\widehat{p}_{a}|W^{\prime }_{(a)},\widehat{p}_{a}^{\prime }; \zeta_{(a)1} \right)
-\tau \left( W_{(a)},\widehat{p}_{a}|W^{\prime }_{(a)},\widehat{p}_{a}^{\prime }; \zeta_{(a)2} \right) \nonumber\\*
=\left( \zeta_{(a)2} - \zeta_{(a)1} \right)
\int \!\!\! \int \, dW^{\prime \prime} \, d \hat{p}^{\prime \prime}
\left( \frac{1}{W^{\prime \prime}-\zeta_{(a)2}} \right)
\left( \frac{1}{W^{\prime \prime}-\zeta_{(a)1}} \right)
\nonumber\\*
\times \tau \left( W_{(a)},\widehat{p}_{a}|W^{\prime \prime}_{(a)},\widehat{p}_{a}^{\prime \prime}; \zeta_{(a)1} \right)
\tau \left( W^{\prime \prime}_{(a)},\widehat{p}_{a}|W^{\prime}_{(a)},\widehat{p}_{a}^{\prime}; \zeta_{(a)2} \right)
\end{eqnarray}

Several relevant outcomes, techniques and definitions have been summarized within this
introduction.  In the next section, parameters and definitions are provided as a foundation
for making a bound state three particle calculation.

\section{\label{sec:level3}Parameterizations and Definitions}

For much of the work that follows, we will use $a$, $a+$,and $a-$ as labels for three bodies,
particles or constituent systems with momenta $k$,$k_+$,$k_-$ in the lab frame.  Parameters $u$ and
$u_{(a)}$ are the Lorentz boost for the CM of the 3 body system and $a+$, $a-$ pair system
respectively.  Choosing to work in the three body CM frame, one may identify body $a$ as
the spectator and define its momentum as $-MU_{(a)}$, and the remaining pair as having momentum $MU_{(a)}$.  In the same
frame, the energies of the respective bodies are $\varepsilon_a$, $\varepsilon_{a+}$, and $\varepsilon_{a-}$.   If $u_{(a)}$ is used to boost to
the pair CM frame, the momenta of the $a+$ and $a-$ particles can be determined, and  are
denoted by  $-P_{(a)}$ and $P_{(a)}$ .  The energies in the pair frame are then,

\begin{eqnarray}
\varepsilon^{(a)}_{a+}
=
\sqrt{m^2_{a+}+|P_{(a)}|^2}
\\*
\varepsilon^{(a)}_{a-}
=
\sqrt{m^2_{a-}+|P_{(a)}|^2}
\end{eqnarray}

The pair invariant energies are then given by the following.

\begin{equation}
M_{(a)}= \sqrt{m^2_{a+} + P^2_{(a)} }+ \sqrt{m^2_{a-} + P^2_{(a)} }= \varepsilon^{(a)}_{a+} + \varepsilon^{(a)}_{a-}
\end{equation}

The variables and parameters just described will now be used to explicitly show the
calculability of LMNP.

\section{\label{sec:level4}Light Particle Model}

In this description of a single light particle model, the particles and systems involved will
be labeled with $a$, $\overline{a}$, and $l$.  We begin by assuming that the masses of $a$, and $\overline{a}$ are
significantly 
greater than $l$ and kinematics of the $a$ and $\overline{a}$ masses.  The intermediate integral of
reference \cite{alfred1} then may be approximated by the following:

\begin{equation}
\Im_{(d)} \left( \underline{u}_{(a)}, \underline{u}_{(d)}^{\prime} , \underline{u}_{(a)} \cdot \underline{u}_{(d)}^{\prime} \right)
\approx
\left[ \frac{\varepsilon^{(a)}_a \varepsilon^{(d)}_d}{P^{\prime}_{(a)} P^{\prime}_{(d)}} \right]^{ \frac{1}{2} }
\frac{ M^2_{(a)} M^2_{(d)} }{ \varepsilon^{\prime}_d \varepsilon^{(d)}_{a} \varepsilon^{(d)}_{d-} u^0_{(a)} }
\end{equation}

The pair energies, momenta and three body energies may also be approximated.

\begin{eqnarray}
M_{(a)} \approx m_{ \overline{a}} \\*
M_{ \overline{a}} \approx m_a \\*
M_{(l)} \approx m_a + m_{ \overline{a}}
\end{eqnarray}

The three body invariant energy may be written in terms of pair energies and velocities.

\begin{equation}
M_{(a)} \approx M_{(a)} u^0_{(a)} + \sqrt{m_a^2 + M^2_{(a)} u^{\prime 2}_{(a)} }
\end{equation}

The pair momenta may be approximated by,

\begin{eqnarray}
\left| P^{(a)}_{ ( \overline{a} ) } \right|^2 & \approx & m^2_{ \overline{a} } u^2_{(a)}
+ m^2_a u^2_{ ( \overline{a} ) } + 2 m_a m_{ \overline{a} } \underline{u}_{(a)} \cdot \underline{u}_{ ( \overline{a} )  } \\
\left| P^{(\overline{a})}_{ ( a ) } \right|^2 & \approx & m^2_{ a } u^2_{(\overline{a})}
+ m^2_{(\overline{a})} u^2_{ ( a ) } + 2 m_{\overline{a}} m_{ a } \underline{u}_{(\overline{a})} \cdot \underline{u}_{ ( a )  }
\\
\left| P^{(l)}_{(a)} \right|^2 & \approx & (m_{ \overline{a} } + m_a)^2 u^2_{(l)}
\end{eqnarray}

If we define bound pair energy, $\mu_{(a)}$ , binding energy $E_B$, and asymptotic spectator
energy parameter, $e_a$, an expression for a separable interaction may be used.  We chose
the following form for the two body scattering amplitude,

\begin{equation}
\tau^{ \ell^{\prime}_a}_a \left( W_{(a)} | W_{(a)}^{\prime} ; \zeta_{(a)}
\right) =
\frac{ 4 \pi g_{(a)} ( W_{(a)} ) g_{(a)}^{\ast} ( W_{(a)}^{\prime} ) \delta_{ \ell 0 }  }{ D_{(a)} ( \zeta_{(a)} ) }
\end{equation}

We chose a simple form for the separable interaction by defining a convenient structure
function for convergence.

\begin{equation}
g(W) = \frac{g_0}{W + \alpha}
\end{equation}

We define the following parameters to provide a direct link between two and three body
space.  The parameters are $\zeta_{(d)}$ , a two body parameter, $e_d$, the spectator asymptotic energy of
a particle with label d, and finally the bound state energy $M_{BS}$.

\begin{eqnarray}
\zeta_{(d)} & = & \frac{(z - e_d)}{u^0_{(d)}}\\
e_d & = & \frac{ M_{BS}^2 - \mu^2_{(d)} + m^2_d  }{2 M_{BS} }\\
M_{BS} & = & m_d + m_{d+} + m_{d-} - E_B
\end{eqnarray}

The two body amplitude $\tau$ has the following form after the unitarity constraint is applied:

\begin{equation}
\frac{ g_{(a)}^{\ast} ( W_{(a)})  g_{(d)} ( W_{(d)}^{\prime}  )   }{ D_{(d)} ( \zeta_{(d)} ) } 
=
\frac{1}{(W^{\prime}_{(a)} + \mu_{(a)}) 4 \pi [F( \zeta_{(d)} ) - F( \mu_{(d)} )] (W^{\prime}_{(d)} + \mu_{(d)})  }
\end{equation}

where the function $F$ may be written in terms of two body space body parameters,

\begin{equation}
F( \zeta_{(d)} ) = \frac{1}{\left(  \mu_{(d)} + \zeta_{(d)}  \right) ^2}
 \left\{ \log \left[ \frac{m_1  + m_2 + \mu_{(d)}}{m_1 + m_2 - \zeta_{(d)}} \right] - \frac{( \mu_{(d)} + \zeta_{(d)} )}{ ( \mu_{(d)} + m_1 + m_2 )  }  \right\}
\end{equation}

\subsection{\label{sec:level4.1}$K_{a \overline{a}}$ Kernel}

For convenience we have the defined the following expressions.

\begin{equation}
A = \frac{1}{u^{0 \prime}} \left[ m_{\overline{a} } u^{0 \prime}_{(a)} + \sqrt{ m^2_a + m^2_{ \overline{a} } u^{2 \prime}_{(a)} } - \alpha
\right]
\end{equation}

The symbol $A$ contains terms constant with respect to the kernel integration.  Having
defined $A$, we may define $b$, $b_1$, $b_2$, $a_1$, and $a_2$ as

\begin{eqnarray}
b & = & b_1 = b_2 = \sqrt{ \frac{A}{m_{ \overline{a} }}  - \frac{ m^2_l  }{ 2 m^2_{ \overline{a}}  }  }
\\
a_1 & = & \frac{ m_l }{ 2b m_{ \overline{a} } }  \left[1 + \sqrt{ 1 - 2b^2 } \right]
\\
a_2 & = & \frac{ m_l }{ 2b m_{ \overline{a} } }  \left[1 - \sqrt{ 1 - 2b^2 }
\right]
\end{eqnarray}

$J_{rel2}$ may now be defined with the above expressions:

\begin{eqnarray}
\left.  J_{rel2}( \alpha )  \right|_{\theta_{-}}^{\theta_{+}}
=
\frac{\mp 2}{u^0_{(a)} m_{\overline{a}} Q_a Q_{\overline{a}} \left( a_2 b_1 - b_2 a_1 \right) }
\nonumber\\
*
\times \left[ \frac{a_2}{\sqrt{b_2^2 - a_2^2} }
\arctan \left( \frac{ \sqrt{b_2^2 - a_2^2} }{a_2 + b_2} \tan \frac{ \theta }{2} \right)
-
\right. \nonumber\\*
\left. \frac{a_1}{\sqrt{b_1^2 - a_1^2} }
\arctan \left( \frac{ \sqrt{b_1^2 - a_1^2} }{a_1 + b_1} \tan \frac{ \theta }{2} \right)
\right]^{\theta_{+}}_{\theta_{-}}
\end{eqnarray}

$Q_a$, and $Q_{ \overline{a} }$ are defined as $Q_a = m_a u_{(a)}$ and $Q_{ \overline{a}}=m_{\overline{a}} u_{( \overline{a} )}$.  After integration, the s wave, $\overline{a} a$ 
channel kernel becomes

\begin{eqnarray}
K^0_{  \overline{a} a } \left(u_{( \overline{a} )} | u^{ \prime }_{(a)} ; z \right)
=
\frac{- \overline{\delta}_{ \overline{a} a} u^{2 \prime}_{(a)} [\varepsilon^{ \overline{a} }_{\overline{a} }  \varepsilon_a^{(a)} ]^{\frac{1}{2}} M^{2 \prime}_{(a)}M^{2 \prime}_{(\overline{a})}
\theta \left( W_{(\overline{a})}^{\prime }-\left( m_{a}+m_{l}\right) \right)
}
{ 2 \pi \varepsilon^{\prime}_a \varepsilon^{(a)}_{\overline{a}} \left[ F \left( z_{(a)} \right) -
F \left( \mu_{(a)} \right) \right]   }
\nonumber\\*
\times \theta \left( W_{(a)}^{\prime }-\left( m_{\overline{a}}+m_{l}\right) \right)
\left[
\frac
{\left.  J_{rel2}( Z )  \right|_{\theta_{-}}^{\theta_{+}}}
{ \left( E^{\prime}_{\overline{a}} - z \right) \left( z - E^{\prime}_a \right) }
+
\right. \nonumber\\*
\left.
\frac
{\left.  J_{rel2}( E^{\prime}_{\overline{a}} )  \right|_{\theta_{-}}^{\theta_{+}}}
{ \left( E^{\prime}_{\overline{a}} - E^{\prime}_a \right) \left( E^{\prime}_{\overline{a}} - z  \right) }
+
\frac
{\left.  J_{rel2}( E^{\prime}_a )  \right|_{\theta_{-}}^{\theta_{+}}}
{ \left( E^{\prime}_{\overline{a}} - E^{\prime}_a \right) \left( E^{\prime}_a  - z \right) }
\right]
\end{eqnarray} 

\subsection{\label{sec:level4.2}$K_{al}$ Kernel}
The $al$ channel and the $\overline{a} l$ channel kernel integrals are found after defining $Q_a = m_a u_{(a)}$
and $Q_l =m_lu_{(l)}$  and a convenient function $B$ is given by

\begin{eqnarray}
B \left( s, \gamma, m \right) & \equiv & \int \frac{dx}{x^{\frac{1}{4}} \left( s x + \gamma   \right)  }
\nonumber\\
& =& \frac{1}{ s^{\frac{3}{4}} \gamma^{ \frac{1}{4} } \sqrt{2} } 
\log \left(
\frac
{1 - \left( \frac{4 \gamma}{s x} \right)^{\frac{1}{4} } + \sqrt{\frac{\gamma x}{s}}   }
{1 + \left( \frac{4 \gamma}{s x} \right)^{\frac{1}{4} } + \sqrt{\frac{\gamma x}{s}}   }
\right)
\nonumber\\
& +
&
\frac{2}{ s^{ \frac{3}{4}} \gamma^{\frac{1}{4}} \sqrt{2} } 
\arctan \left(
\frac{ \left( 4 s \gamma x \right)^{\frac{1}{4}} }
{\gamma^{\frac{1}{2}} - \left( s x \right)^{\frac{1}{2} }} 
\right)
\end{eqnarray}

Using the above definitions, the S-wave $al$ kernel may be written as the following.

\begin{eqnarray}
K^0_{ al } \left(u_{(a)} | u^{ \prime }_{(l)} ; z \right)
=
\frac{
- \overline{\delta}_{  a l} u^{2 \prime}_{(l)} [\varepsilon_l^{(l) }  \varepsilon_a^{(a)} ]^{\frac{1}{2}} M^{2 \prime}_{(l)}M^{2 \prime}_{(a)}
\theta \left( W_{(a)}^{\prime }-\left( m_{\overline{a}}+m_{l}\right) \right)
}
{ 2 \pi \varepsilon^{\prime}_l \varepsilon^{(l)}_a \varepsilon^{(l)}_{(\overline{a})}
\left[ P^{\prime}_{(a)} \right]^{\frac{1}{2}} \left[ F \left( \zeta_{(l)} \right) - F \left( \mu_{(l)} \right)  \right]  }
\nonumber\\*
\times \theta \left( W_{(l)}^{\prime }-\left( m_a + m_{\overline{a}}\right) \right)
\frac{ m_a^{\frac{1}{2}} }{ 2 Q_a Q_l }
\left[ 
\frac
{B \left( \frac{ u^{0 \prime}_{(l)} m_a}{2 n_{(l)}}, \frac{C}{m_a},z  \right)}
{\left( E^{\prime}_a - z \right) \left( z - E^{\prime}_l \right)  }
\right. \nonumber\\*
\left.
+
\frac
{B \left( \frac{ u^{0 \prime}_{(l)} m_a}{2 n_{(l)}}, \frac{C}{m_a},E_a^{\prime}  \right)}
{\left( E^{\prime}_a - E^{\prime}_l \right) \left( E^{\prime}_a - z \right)}
+
\frac
{B \left( \frac{ u^{0 \prime}_{(l)} m_a}{2 n_{(l)}}, \frac{C}{m_a},E_l^{\prime} \right) }
{\left( E^{\prime}_a - E^{\prime}_l \right) \left( E^{\prime}_l - z \right)}
\right]_{\frac{{Q_min}}{m_a} }^{\frac{Q_{max}}{m_a}}
\end{eqnarray}

We have combined many parameters and variables, constant with respect to the
integration of the $K_{al}$ kernel, into the constant $C$.

\begin{equation}
C= (m_a + m_{\overline{a}}) u^{0 \prime}_{(l)}+\sqrt{m^2_l + (m_a + m_{\overline{a}}^2) u^{\prime 2}_{(l)} } - \alpha
\end{equation}

\subsection{\label{sec:level4.3}$K_{la}$ Kernel}
After integration, the $K_{la}$ S-wave kernel may be written as the following:

\begin{eqnarray}
K^0_{ la } \left(u_{(l)} | u^{ \prime }_{(a)} ; z \right)
=\frac
{
- \overline{\delta}_{  l a} u^{2 \prime}_{(a)} [\varepsilon_l^{(l) }  \varepsilon_a^{(a)} ]^{\frac{1}{2}} M^{2 \prime}_{(l)}M^{2 \prime}_{(a)}
\theta \left( W_{(a)}^{\prime }-\left( m_{\overline{a}}+m_{l}\right) \right)
}
{ 2 \pi \varepsilon^{\prime}_a \varepsilon^{(a)}_l \varepsilon^{(a)}_{(\overline{a})}
\left[ P^{\prime}_{(a)} \right]^{\frac{1}{2}} \left[ F \left( \zeta_{(a)} \right) - F \left( \mu_{(a)} \right) \right]    }
\nonumber\\*
\times \theta \left( W_{(l)}^{\prime }-\left( m_a + m_{\overline{a}}\right) \right)
\left[
\frac{ m_a^{\frac{3}{2}} x^{\frac{3}{4}} }
{ \left( M^{a \prime} - z \right) \left( M^{a \prime} - E^{\prime}_l \right) \left( M^{a \prime} - E^{\prime}_a \right) Q_a Q_l   }
\right]_{x=\frac{Q^2_{min}}{m_a^2}}^{x=\frac{Q^2_{max}}{m_a^2}}
\end{eqnarray}

For convenience, we have defined $Q_a=m_a u_{(a)}$, $Q_l=m_l u_{(l)}$.  $Q_{max}$ and $Q_{min}$ are the
maximum and minimum values of the addition of $Q_a$ and $Q_l$.
     
\section{\label{sec:level5}Ultra-light Quantum Extension}

The light model kernels have been described in the previous section.  We will now
present an ultra-light extension of the light model.  After stating required assumptions,
approximations are made yielding a single kernel channel.  The final form of the extended
model is presented with a modified scaling term for reasonable numbers during
calculation.

To extend the light model, an assumption is made that the light particle alone mediates
the interaction between two equally massive particles; only the $a \overline{a}$ channel exists.  In
this regard it serves as the quantum of that interaction.  The assumptions made of the
single channel and relativistic energies allow strict approximations to be made of
spectator energies.

\begin{eqnarray}
\varepsilon^{ \left( \overline{a} \right)}_{ \overline{a} }
& = &
u^0_{ \left( \overline{a} \right) }
\sqrt{ m^2_{ \left( \overline{a} \right) } + \left( m_a u^{ \prime }_{ \left( \overline{a} \right) } \right)^2 } 
+
m_a u^{\prime 2}_{ \left( \overline{a} \right) } \approx m_{ \left( \overline{a} \right) }
\\
\varepsilon^{(a)}_a 
& = &
u^0_{(a)} \sqrt{ m^2_a + \left( m_{ \left( \overline{a} \right) } u_{(a)}^{ \prime } \right)^2}
+ m_{\overline{a}} u_{(a)}^{'2}
\approx m_a
\\
\varepsilon^{ \prime }_{ \left( \overline{a} \right) } 
& = &
\sqrt{m^2_{  \overline{a} } + \left(  m_a u^{ \prime }_{ \left( \overline{a} \right) } \right)^2 } \approx m_{ \left( \overline{a} \right) }
\\
\varepsilon_a^{ \prime \left( \overline{a} \right) }
& = &
\sqrt{ m^2_a + P^2_{ \left( \overline{a} \right) } } \approx m_a
\\
M_{(a)} & \approx & m
\\
M_{ \left( \overline{a} \right) } & \approx & m
\end{eqnarray}

A separable interaction which takes the same form as that of the light model
approximation, is assumed for the ultra light extension.  In addition, the $F$ function is
approximated in exactly the same way as was done in the light model.  The inclusion of
all approximations and assumptions for the ultra-light model yields a single two
dimensional integral for the kernel,

\begin{eqnarray}
K^L_{  \overline{a} a } \left(u_{( \overline{a} )} | u^{ \prime }_{(a)} ; z \right)
=\frac
{
- \overline{\delta}_{ \overline{a} a} u^{\prime 2}_{(a)} 2 \pi [\varepsilon^{ \overline{a} }_{\overline{a} }  \varepsilon_a^{(a)} ]^{\frac{1}{2}} M^{\prime 2}_{(a)}M^{\prime 2}_{(\overline{a})}
\theta \left( W_{(\overline{a})}^{\prime }-\left( m_{a}+m_{l}\right) \right)
}
{ u^{0 \prime}_{\overline{a}} u^0_{(a)} \varepsilon^{\prime}_a \varepsilon^{(a)}_{\overline{a}} 
4 \pi \left[ F \left( z_{(a)} \right) - F \left( \mu_{(a)} \right) \right]   }
\nonumber\\*
\times \theta \left( W_{(a)}^{\prime }-\left( m_{\overline{a}}+m_{l}\right) \right)
\int_{-1}^1
\frac
{P_L \left( \hat{u}_{\left( \overline{a} \right)  } \cdot \hat{u}^{ \prime }_{(a)} \right) \, d \left( \hat{u}_{\left( \overline{a} \right)  } \cdot \hat{u}^{ \prime }_{(a)} \right)}
{ \left[ P^{\prime}_{(a)} \right] \varepsilon^{(a)}_l
\left( M^{ \prime } - z \right) \left( W_{( \overline{a})} + \mu_{( \overline{a} )} \right)
\left( W_{(a)} + \mu_{(a)} \right)
}
\end{eqnarray}

The kernel may be simplified with the definition of a binding energy $E_B$ and constant
parameter $H$.

\begin{equation}
z = M_{BS} = m_a + m_{ \overline{a} } + m_l - E_B
\end{equation}

\begin{equation}
H \equiv \frac{1}{u^0_{( \overline{a} )}} \left( m_a \left( u^0_{ \overline{a} } - 1 \right) + \left( \sqrt{ m^2_{ \overline{a} } + m^2_a u^{ \prime 2}_{ ( \overline{a} ) }  } - m_{ \overline{a} } - m_l + E_B \right) \right)
\end{equation}

As with the light quantum model, it is useful to define parameters $Q_a$, and $Q_{\overline{a}}$ for the
limits of the integration.  The final expression for the integrated kernel may then be
written as the following:

\begin{eqnarray}
\lim_{ULQ} K_{  \overline{a} a } \left(u_{( \overline{a} )} | u^{ \prime }_{(a)} ; z \right)
\approx
\frac
{- \overline{\delta}_{ \overline{a} a} u^{2 \prime}_{(a)} 2 \pi [\varepsilon^{ \overline{a} }_{\overline{a} }  \varepsilon_a^{(a)} ]^{\frac{1}{2}} M^{2 \prime}_{(a)}M^{2 \prime}_{(\overline{a})}
}
{ u^{0 \prime}_{\overline{a}} u^0_{(a)} \varepsilon^{\prime}_a \varepsilon^{(a)}_{\overline{a}} 
4 \pi \left[ F \left( z_{(a)} \right) - F \left( \mu_{(a)} \right) \right]   }
\nonumber\\*
\times \frac{\theta \left( W_{(\overline{a})}^{\prime }-\left( m_{a}+m_{l}\right) \right)
\theta \left( W_{(a)}^{\prime }-\left( m_{\overline{a}}+m_{l}\right) \right)
}
{\left( W_{( \overline{a})} + \mu_{( \overline{a} )} \right) \left( W_{(a)} + \mu_{(a)} \right)
m_a u^0_{( \overline{a} )} Q_a Q_{ \overline{a} } 
\sqrt{ \left( \frac{H}{m_a} \right)^2 - \left( \frac{m_l}{m_a} \right)^2}     } \nonumber\\*
\times \log 
\left|
\frac
{ \sqrt{H-m_l}P + \sqrt{H+m_l} \left( \varepsilon + m_l \right)   }
{ \sqrt{H-m_l}P - \sqrt{H+m_l} \left( \varepsilon + m_l \right)   }
\right|_{P_+ = | Q_a + Q_{ \overline{a} } | }^{P_- = | Q_a - Q_{ \overline{a} } | }
\end{eqnarray}

It is now desirable to scale the kernel so that its values are of order unity.  For this reason,
two parameters will be defined for calculational purposes.  We define  
 
\begin{equation}
G \left( \tilde{u}^{ \prime }, \tilde{u}, m_l, \mu, ES \right) = 
\frac{1}{\left( F \left( z_{( \overline{a} )} \right) - F \left( \mu_{( \overline{a} )} \right) \right) \left( W_{( \overline{a} )} + \mu_{( \overline{a} )} \right) \left( W_{(a)} + \mu_{(a)} \right)
}
\end{equation}
 
The constant beta is then defined in terms of $G$, 
 
\begin{equation}
\beta = G \left( 0, 0, m_l, m_l, 0 \right)
\end{equation}
 
We may now rescale $u$ and $H$ in the following way: 
 
\begin{eqnarray}
H & = & \beta^2 \tilde{H} \\
u & = & \beta \tilde{u}
\end{eqnarray}
 
A final spin zero kernel that we will use for bound state energy calculations may now be
explicitly written:

\begin{eqnarray}
K^0_{  \overline{a} a } \left(u_{( \overline{a} )} | u^{ \prime }_{(a)} ; z \right)
& \approx
&
\frac
{ \tilde{u}^{\prime}_{(\overline{a})} G \left( \beta \tilde{u}^{ \prime }_{(a)}, \beta \tilde{u}_{( \overline{a} )}, m_l, \mu, ES \right) }
{\tilde{u}_{(a)} \beta^2 H}
\nonumber\\*
& & \times \log
\left|
\frac
{\sqrt{ \tilde{H}+ \frac{m_l}{\beta^2} } \left( \tilde{\varepsilon}_l + \frac{m_l}{\beta} \right) + \sqrt{ \tilde{H} - \frac{m_l}{\beta^2} } \tilde{P}  }
{\sqrt{ \tilde{H}+ \frac{m_l}{\beta^2} } \left( \tilde{\varepsilon}_l + \frac{m_l}{\beta} \right) - \sqrt{ \tilde{H} - \frac{m_l}{\beta^2} } \tilde{P}  }
\right|
\end{eqnarray}

\section{\label{sec:level6}Results}

The primary goal of this work has been to attempt a calculation of bound state energies 
using the LMNP formalism, explicitly demonstrating calculability.  The direct form of the
ultra-light kernel mentioned above was solved for bound state energies using gaussian
quadrature and Newton methods for eigen-integral equations.  

\begin{table}
\caption{\label{tab:table1} Energy results}
\begin{ruledtabular}
\begin{tabular}{lcr}
Bound state & Energy/(m) & Uncertainty(energy/m)\\
\hline
1 & .000030  & $8 \times 10^{-6}$ \\
2 & .000020 &  $8 \times 10^{-6}$ \\
3 & .000015  &  $8 \times 10^{-6}$ \\
4 & .000015  &  $6 \times 10^{-6}$ \\
5 & .000013  &  $6 \times 10^{-6}$ \\
6 & .000012 & $7 \times 10^{-6}$ \\
\end{tabular}
\end{ruledtabular}
\end{table}

As listed in Table~\ref{tab:table1}, six stable bound state energies were found as the ultra-light quantum mass $m_l$ approached zero.

\begin{figure}
	\includegraphics{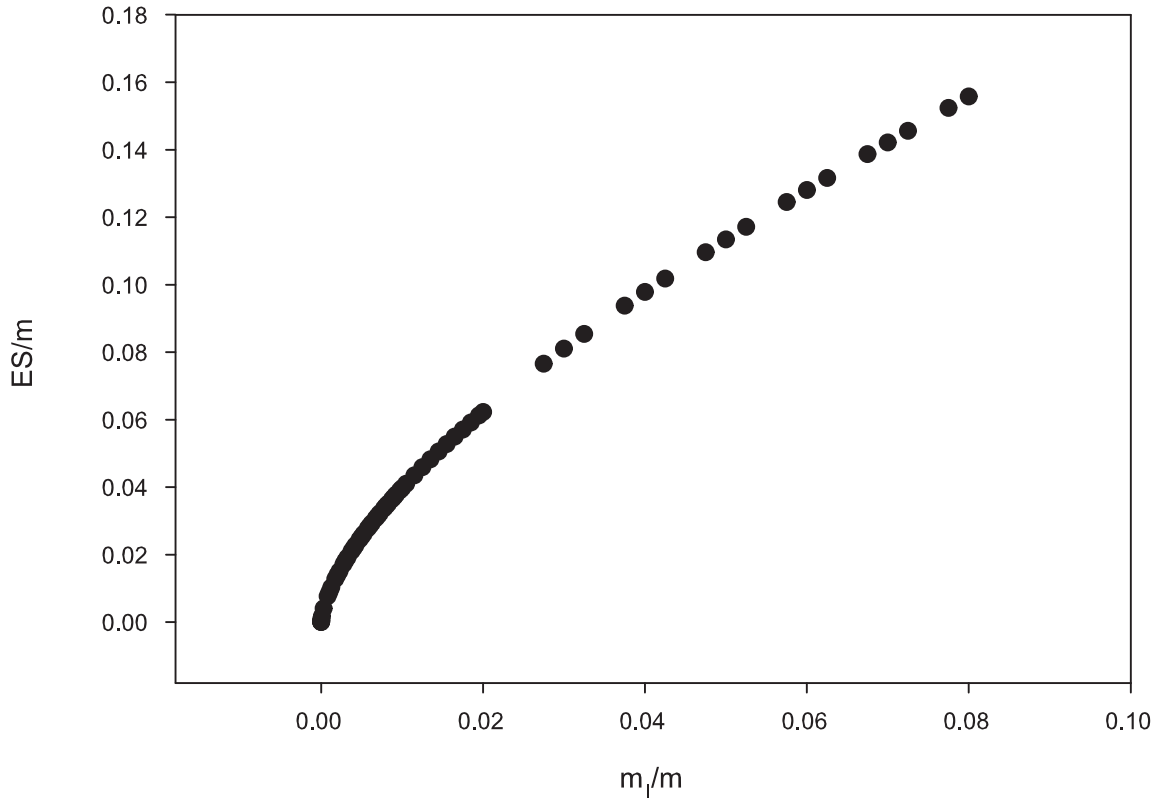}
	\caption{\label{fig1}Depicted is the 1s bound state energy versus the quantum mass $m_l$, both scaled by the massive particle mass m.}
\end{figure}

The nonzero asymptotic energy of the first bound state, $ES = E_B -m_l$,  is presented in
Fig.~\ref{fig1} as the ultra-light quantum mass $m_l$ approaches zero.

\section{\label{sec:level7}Conclusion}

We have demonstrated the calculability of the LMNP formalism, by constructing both a simple light particle model and its ultra-light quantum
extension, and by obtaining bound state energies for the latter.  After modifications, the non-relativistic
parameterizations and assumptions of both models may be used for other bound state calculations.  In particular, by 
using alternative two body amplitudes in the kernel expressions of the light model, one can model a wide range of 
nuclear and particle interactions.  In summary, the ultra-light quantum and light models, in addition to their
associated methods, may work as effective templates for general bound state problems.
         
\begin{acknowledgments}
M.A. would like to thank the Howard University CSTEA, MGE, GAANN, and Dissertation Fellowship programs for their generous support during the development of this work.
\end{acknowledgments}

\bibliography{scat2} 

\end{document}